\documentclass[twoside, twocolumn, 10pt]{article}
\usepackage{extsizes}
\usepackage[super, sort&compress, comma]{natbib}

\usepackage[version=3]{mhchem}
\usepackage[left=1.5cm, right=1.5cm, top=1.785cm, bottom=2.0cm]{geometry}
\usepackage{balance}
\usepackage{mathptmx}
\usepackage{sectsty}
\usepackage{graphicx}
\usepackage{lastpage}

\usepackage[
    format=plain,
    justification=justified,
    singlelinecheck=false,
    font={stretch=1.125,small,sf},
    labelfont=bf,
    labelsep=space
]{caption}
\usepackage{float}
\usepackage{fancyhdr}
\usepackage{fnpos}
\usepackage[english]{babel}
\addto{\captionsenglish}{%
 }
\usepackage{array}
\usepackage{droidsans}
\usepackage{charter}
\usepackage[T1]{fontenc}
\usepackage[usenames, dvipsnames]{xcolor}
\usepackage{setspace}
\usepackage[compact]{titlesec}
\usepackage[hidelinks]{hyperref}
\usepackage[mathscr]{euscript}
\usepackage{makecell}
\usepackage{color}

\ExplSyntaxOn
\NewDocumentCommand{\MeijerG}{smmmm}{ \IfBooleanTF{#1}
{ \vic_meijerg:nnnnnn { #2 } { #3 } { #4 } { #5 } { small } { } }
{ \vic_meijerg:nnnnnn { #2 } { #3 } { #4 } { #5 } { } { \; } } }

\seq_new:N
\l__vic_meijerg_args_in_seq
\seq_new:N
\l__vic_meijerg_args_out_seq

\cs_new_protected:Nn
\vic_meijerg:nnnnnn
{ \seq_set_split:Nnn \l__vic_meijerg_args_in_seq { | } { #3 } \seq_clear:N \l__vic_meijerg_args_out_seq \seq_map_inline:Nn \l__vic_meijerg_args_in_seq { \seq_put_right:Nn \l__vic_meijerg_args_out_seq { \begin{#5matrix}##1\end{#5matrix} } } G\sp{#1}\sb{#2} \left( \seq_use:Nn \l__vic_meijerg_args_out_seq { #6\middle|#6 } #6\middle|#6 #4 \right) }
\ExplSyntaxOff

\newcommand{\Det}{\mathrm{Det}}
\usepackage{nicematrix}
\newcommand{\I}{\mathrm{I}}

\newcommand{\bea}{\begin{eqnarray}}
\newcommand{\eea}{\end{eqnarray}}
\usepackage{hyperref}
\hypersetup{
    colorlinks=true,
    urlcolor=blue,
    citecolor=blue, 
    linkcolor=magenta 
}

\usepackage{soul}

\usepackage{amsmath}
\usepackage{amssymb}
\DeclareRobustCommand\hbar{{\mathchar'26\mkern-9muh}}

\usepackage{
    epstopdf
}

\usepackage{multirow} 

\global\arraycolsep=2pt
\usepackage{stmaryrd}
\usepackage{amsmath}
\usepackage{amssymb}
\usepackage{graphicx}
\usepackage{textcomp}
\usepackage{calrsfs}
\usepackage{caption}
\usepackage{subcaption}
\usepackage{yfonts}
\usepackage{bm}
\usepackage{soul}
\usepackage{orcidlink}
\usepackage{color}
\newcommand{\ben}{\begin{equation*}}
\newcommand{\een}{\end{equation*}}
\newcommand{\bean}{\begin{eqnarray*}}
\newcommand{\eean}{\end{eqnarray*}}

\newcommand{\be}{\begin{equation}}
\newcommand{\ee}{\end{equation}}

\definecolor{cream}{RGB}{222,217,201}

\usepackage{soul}

\begin{document}
    \pagestyle{fancy}
    \thispagestyle{plain}
    \fancypagestyle{plain}{
    \renewcommand{\headrulewidth}{0pt} }

    \makeFNbottom \makeatletter
    \renewcommand{\LARGE}{\@setfontsize\LARGE{15pt}{17}}
    \renewcommand{\Large}{\@setfontsize\Large{12pt}{14}}
    \renewcommand{\large}{\@setfontsize\large{10pt}{12}}
    \renewcommand{\footnotesize}{\@setfontsize\footnotesize{7pt}{10}}
    \makeatother

    \renewcommand{\thefootnote}{\fnsymbol{footnote}}
    \renewcommand{\footnoterule}{\vspace*{1pt}%
    \color{cream}
    \hrule width 3.5in height 0.4pt
    \color{black}
    \vspace*{5pt}}
    \setcounter{secnumdepth}{5}

    \makeatletter
    \renewcommand{\@biblabel}[1]{#1}
    \renewcommand{\@makefntext}[1]%
    {\noindent\makebox[0pt][r]{\@thefnmark\,}#1} \makeatother
    \renewcommand{\figurename}{\small{Fig.}~}
    \sectionfont{\sffamily\Large} \subsectionfont{\normalsize} \subsubsectionfont{\bf}
    \setstretch{1.125} 
    \setlength{\skip\footins}{0.8cm}
    \setlength{\footnotesep}{0.25cm}
    \setlength{\jot}{10pt}
    \titlespacing*{\section}{0pt}{4pt}{4pt} \titlespacing*{\subsection}{0pt}{15pt}{1pt}

    \fancyfoot{} \fancyfoot[LO,RE]{\vspace{-7.1pt}\includegraphics[height=9pt]{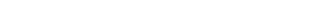}}
    \fancyfoot[CO]{\vspace{-7.1pt}\hspace{11.9cm}\includegraphics{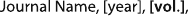}}
    \fancyfoot[CE]{\vspace{-7.2pt}\hspace{-13.2cm}\includegraphics{head_foot/RF}}
    \fancyfoot[RO]{\footnotesize{\sffamily{1--\pageref{LastPage} ~\textbar \hspace{2pt}\thepage}}}
    \fancyfoot[LE]{\footnotesize{\sffamily{\thepage~\textbar\hspace{4.65cm} 1--\pageref{LastPage}}}}
    \fancyhead{}
    \renewcommand{\headrulewidth}{0pt}
    \renewcommand{\footrulewidth}{0pt}
    \setlength{\arrayrulewidth}{1pt}
    \setlength{\columnsep}{6.5mm}
    \setlength{\bibsep}{1pt}

    \makeatletter
    \newlength{\figrulesep}
    \setlength{\figrulesep}{0.5\textfloatsep}

    \newcommand{\topfigrule}{\vspace*{-1pt}%
    \noindent
    {\color{cream}\rule[-\figrulesep]{\columnwidth}{1.5pt}} }

    \newcommand{\botfigrule}{\vspace*{-2pt}%
    \noindent
    {\color{cream}\rule[\figrulesep]{\columnwidth}{1.5pt}} }

    \newcommand{\dblfigrule}{\vspace*{-1pt}%
    \noindent
    {\color{cream}\rule[-\figrulesep]{\textwidth}{1.5pt}} }

    \makeatother

    \twocolumn[
    \begin{@twocolumnfalse}
        {\includegraphics[height=30pt]{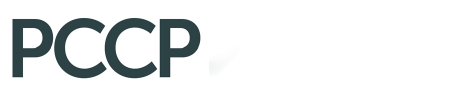}\hfill\raisebox{0pt}[0pt][0pt]{\includegraphics[height=55pt]{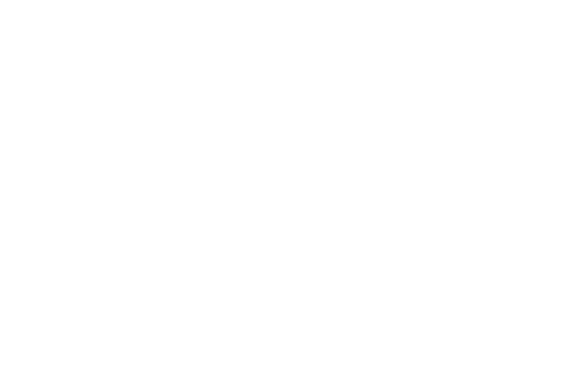}}\\[1ex] \includegraphics[width=18.5cm]{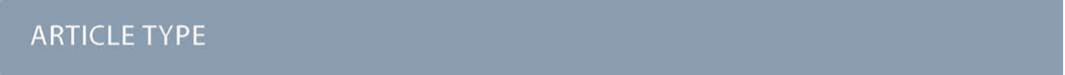}}\par
        \vspace{1em}
        \sffamily
        \begin{tabular}{m{4.5cm} p{13.5cm} }
            \includegraphics{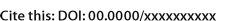}   & \noindent\LARGE{\textbf{Attractive and repulsive terms in multi-filament dispersion interactions}}                                                                                                                                                                                                                                                                                                                                                                                                                                                                                                                                                                                                                                                                                                                                                                                                                                                                                                                                                                                                                                                                                                                                                                                                                                                                                                                                                                     \\ 
            \vspace{0.3cm}                    & \vspace{0.3cm}                                                                                                                                                                                                                                                                                                                                                                                                                                                                                                                                                                                                                                                                                                                                                                                                                                                                                                                                                                                                                                                                                                                                                                                                                                                                                                                                                                                                                                                         \\
                                              & \noindent\large{Subhojit Pal\,\orcidlink{0009-0003-6356-3409},$^{\ast}$\textit{$^{a}$}, John F. Dobson\,\orcidlink{0000-0002-7582-1378}\,$^{\ast}$,\textit{$^{b}$}, and Mathias Bostr\"om\,\orcidlink{0000-0001-5111-4049}\,$^{\ast}$\textit{$^{c,d}$}}                                                                                                                                                                                                                                                                                                                                                                                                                                                                                                                                                                                                                                                                                                                                                                                                                                                                                                                                                                                                                                                                                                                                                                                                                \\
            \includegraphics{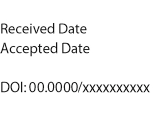} & \noindent\normalsize{ Filamentary objects such as nano-wires, nanotubes and DNA are of current interest in physics, nanoscience, chemistry, biology and medicine. They can interact via strong, exceptionally long-ranged many-object van der Waals (vdW, dispersion) forces, causing them to cluster into multi-object bundles. We analyse their vdW interactions perturbatively, predicting $N$-object vdW energy contributions that alternate in sign with increasing $N$. Our findings are confirmed here via the first detailed analysis of a 4-cylinder vdW model. We also provide novel insights permitting these tendencies to be understood simply in terms of electronic screening and anti- screening. Our results suggest that a non-perturbative calculation will be required for reliable prediction of dispersion interactions in these ubiquitous systems.
} \\ 
        \end{tabular}
    \end{@twocolumnfalse}
    \vspace{0.6cm}
    ]

    \renewcommand*{\rmdefault}{bch}
    \normalfont\upshape \rmfamily
    \section*{}
    \vspace{-1cm}


    \footnotetext{\textit{$^{a}$~Dipartimento di Fisica e Chimica, Emilio Segr\`e,
    \href{https://www.unipa.it/target/international-students/en/about/the-university/}{Universit\`a
    degli Studi di Palermo}, Via Archirafi 36, 90123 Palermo, Italy; Email: \href{palsubhojit429@gmail.com}{palsubhojit429@gmail.com}}}

    \footnotetext{\textit{$^{b}$~School of Environment and Science, \href{https://www.griffith.edu.au}{Griffith University},
    Nathan, Queensland 4111, Australia; Email:
    \href{john12dobson@gmail.com}{john12dobson@gmail.com} }}

    \footnotetext{\textit{$^{c}$~\href{https://ensemble3.eu/}{Centre of
    Excellence ENSEMBLE3}
    Sp. z o. o., Wolczynska Str. 133, 01-919, Warsaw, Poland; E-mail:
    \href{mathias.bostrom@ensemble3.eu}{mathias.bostrom@ensemble3.eu}}}
    \footnotetext{\textit{$^{d}$~Chemical and Biological Systems Simulation Lab,
    \href{https://cent.uw.edu.pl/}{Centre of New Technologies, University of Warsaw},
    Banacha 2C, 02-097 Warsaw, Poland. }}

    \footnotetext{\textit{$^{\ast}$~Corresponding Authors}}





    \section{Introduction}

    Dispersion (van der Waals, vdW) forces are strongly affected by the shape of
    the interacting objects, and thin elongated (filamentary) structures are a particularly
    interesting case. Filamentary structures include nanowires, nanotubes and DNA,
    and they occur very widely in colloid science\,\cite{Ninhb} and biology\,\cite{2strandedDNA,3strandedDNA,4strandedDNA,reines_ninham_2019,ninham2023nafion,NINHAM2025103401}.
    For example, it was recently pointed out\,\cite{ninham2023nafion} that the
    endothelium of all living cells contains such filaments. The electrical polarizability
    of filaments tends to be much higher longitudinally than transversely, and
    this is especially so because many examples such as metal nanowires, (n,n) carbon
    nanotubes, and even DNA are highly conductive, with other biological examples
    also believed to be conductors\,\cite{NINHAM2025103401,chang1971van}. This
    high polarizability makes the vdW interaction between filaments unusually
    strong. In the metallic cases the vdW interaction E(R) between a pair of
    filaments is also known\,\cite{langbein1972van,mitchell1973van,SmithJTheoPhys1973,Maha,Pars,dobson2006asymptotics,misquitta2010dispersion,Richmond1972,drummond2007van}
    to fall off extremely slowly with increasing separation $R$: $E\propto -R^{-2}$
    or $-R^{-1}$ for metallic filaments whereas $E\propto -R^{-5}$ for insulating
    filaments.

    The simplest theories of vdW interactions add up two-atom (or two-object) vdW
    energy contributions\,\cite{BarkerHendersson1967,Barker1971LiquidAM}.
    Following the work of Axilrod, Teller and Muto\,\cite{AxilrodTeller1943,muto1943force},
    however, it was realised that three-object terms could be important\,\cite{RichardsonMahanty_1977,Anatole2010,PhysRevB.68.205409}.
    For example, the crystal structure of rare-gas solids is determined by the 3-atom
    vdW interaction\,\cite{Maha}. vdW energy contributions beyond triplets can
    also occur. As will be explained below, such multi-object contributions
    arise from Coulomb screening (or anti-screening) effects caused by the
    addition of further polarizable objects to an existing cluster. Therefore,
    since filamentary objects are highly polarizable, they experience strong beyond-pairwise
    vdW interactions. Previous studied of 3-object vdW forces have shown that
    the sign (attractive or repulsive) of beyond-pairwise contributions is
    highly dependent on shape (e.g. spherical or elongated) and geometrical arrangement
    (e.g. in a collinear array or an equilateral triangle array). Thus detailed numerical
    calculations are required to determine the sign of beyond-pairwise energy contributions.
    This can be understood as arising from a delicate competition between
    screening and anti-screening. \\

    An essential point of this present work is that detailed calculation is not
    required to determine these signs for the widely occurring case of $N$
    parallel elongated objects that are primarily polarizable longitudinally. This
    is because anti-screening does not occur. As a result, the 3-object vdW energy
    term is always positive (repulsive), and we will prove for the first time
    here that the sign of the leading $N$-object term is $(-1)^{N+1}$.

    The sign alternation means that the level of convergence will be hard to
    determine from a small number of terms in a perturbative expansion of the
    vdW energy. Thus a non-perturbative calculation will be required, which is
    another significant conclusion of the present work. \\ The paper is organized
    as follows. In Section\,(\ref{sec2}) we discuss the concepts of screening
    and anti-screening as applied to many-object vdW interactions. In Section\,(\ref{sec3})
    we quantitatively analyse the $N$-object term in a perturbative expansion of
    the vdW interaction between $N$ quasi-one-dimensional objects. We prove that
    the sign of this term is $(-1)^{N+1}$. In Section\,(\ref{sec4}) we confirm the
    above results for $N$=4 within a plasma cylinder model that goes beyond the atomically-thin
    limit of quasi-one-dimensional objects. In Section\,(\ref{BiologyNano}) we
    discuss the implications of our results for accurate modeling of interacting
    filamentary objects which, as explained above, are ubiquitous in nano- and bio-physical
    situations.

    \section{Qualitative argument: Screening effects}
    \label{sec2} In general, dispersion interactions can be understood
    conceptually as the time-averaged Coulomb energy between a spontaneous
    multipole fluctuation on one object, and the multipole that it induces on another
    object (see eg. Ch.\,2 of Ref.\cite{angyan2020london}). The occurrence of
    beyond-pairwise vdW contributions is sometimes termed ``type-B nonadditivity''\,\cite{dobson2014}.
    From this viewpoint the $N$-object term can be attributed to the screening (or
    anti-screening) of the vdW interactions among $N-1$ objects, due to the
    introduction of an additional $N^{\mathrm{th}}$ polarizable object. We will term
    such $N$-object contributions ``irreducible'' meaning that they cannot be expressed
    as a sum of $m$-object terms where $m<N$. \\

\begin{figure}[!htb]
        \centering
        \includegraphics[width=0.7\linewidth]{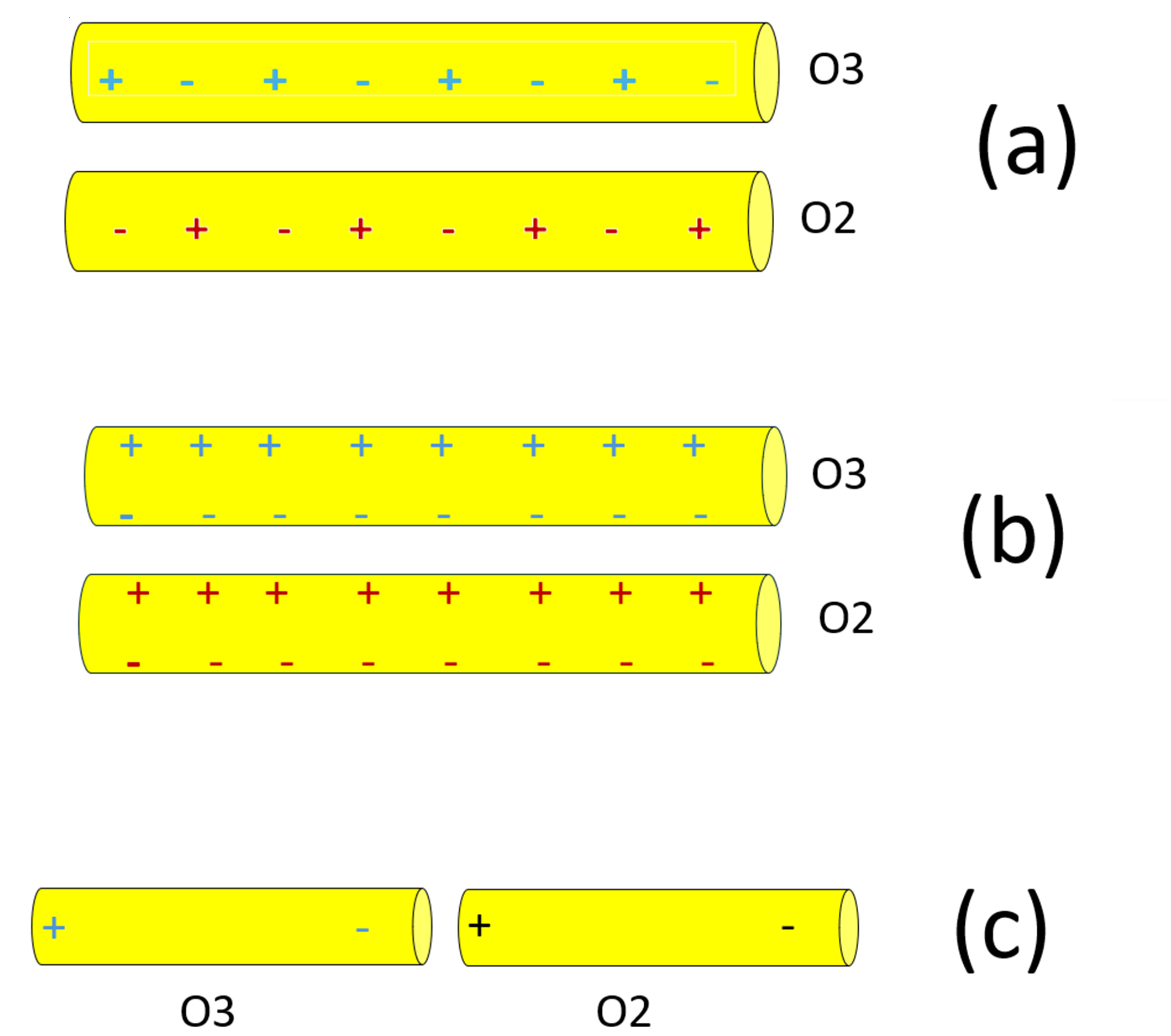}
        \caption{(Colors online) screening and anti-screening for elongated
        objects. (a) Screening for longitudinally polarizable parallel objects.
        A multipole on object O2 Coulomb-induces a contrary multipole on O3 (faint
        ``$\pm$ '' symbols) (b) Anti-screening (enhancement) occurs when the polarizability
        is predominantly in the direction joining the objects. Here a dipole on
        O2 induces a similar dipole on O3. (c) Anti-screening when longitudinally
        polarizable objects are collinear. }
        \label{fig:one}
    \end{figure}

    Screening and anti-screening are illustrated in the present context by Fig.\,(\ref{fig:one})
    depicting two elongated objects.
    It shows the polarization (solid ``$\pm$''symbols) that has been Coulomb-induced
    object O2 by a spontaneous multipole fluctuation on an object O1. (O1 is not
    shown in the diagram). The polarization on O2 then Coulomb-induces a
    polarization on O3 (faint ``$\pm$'' symbols). Fig.\,(\ref{fig:one}a) corresponds
    to the objects of primary interest here, which are longitudinally
    polarizable. Here we have screening, meaning that induced charge
    distribution on O3 is opposite to that on O2. Thus the combined system (O2+O3)
    has its longitudinal polarizability $\alpha$ reduced:
    $0<\alpha^{(O2+O3)}<\alpha^{(O2)}+\alpha^{(O3)}$. By contrast, when the
    objects are polarizable primarily along the x-axis pointing between the
    parallel objects as in Fig.\,(\ref{fig:one}b), we have anti-screening: the
    induced multipole on O3 reinforces that on O2, so that the combined polarizability
    is enhanced: $\alpha^{(O2+O3)}>\alpha^{(O2)}+\alpha^{(O3)}>0$. We can understand
    the effect of this screening phenomenon on the 3-object vdW interaction as follows.
    A well-known argument (see e.g. Ch.\,2 of Ref\,\cite{angyan2020london})
    based on the above-mentioned ``spontaneous + induced multipole" concept shows
    that, at fixed spatial separation, the vdW energy $g^{(a,b)}$ between two objects
    $a$ and $b$ is proportional to the polarizability product:
    $g^{(a,b)}\propto - \alpha^{(a)}\alpha^{(b)}$, or more precisely to the frequency
    integral of this product. When a third object O3 is introduced to a pair O1,
    O2, the vdW energy is the sum of a new pair interaction $g^{(O2,O3)}$ and
    the pair interaction between O1 and the new combined object (O2+O3): $g^{(O1+O2+O3)}
    = g^{(O2,O3)}+ g^{(O1,(O2+O3))}$. For the ``screening'' geometry (Fig.\,(\ref{fig:one}a))
    the above polarization inequality shows that
    $g^{(O1,(O2+O3))}\propto - \alpha^{(O1)}\alpha^{(O2+O3)}> - \alpha^{(O1)}\Big
    (\alpha^{(O2)}+ \alpha^{(O3)}\Big)$. The total interaction is thus reduced in
    magnitude (is less negative) compared with the sum of pairwise energies:
    $0 > g^{(O1+O2+O3)}> g^{(O2,O3)}+ g^{(O1,O2)}+ g^{(O1,O3)}$. This amounts to
    an irreducible 3-object energy that is positive (repulsive). In comparison, for
    the ``anti-screening'' geometry of Fig.\,(\ref{fig:one}b), the
    polarizability inequality is reversed so the irreducible 3-object energy is
    negative (attractive). Fig.\,(\ref{fig:one}c) shows another ``anti-screening''
    geometry that yields negative (attractive) 3-object energy term. The above
    argument needs to be symmetrized with respect to the object labels, but is plausible,
    nevertheless. It might be generalizable, suggesting sign alternation with
    increasing number $N$ of objects.

    \section{Quantitative arguments for attractive/repulsive interactions in the
    N-object case}
    \label{sec3}

    Below we will sketch a more general derivation with just enough detail to establish
    the sign of the irreducible $N$-object contribution to the dispersion energy
    of $N$ disjoint parallel uniaxially polarizable linear objects. Let
    $\chi^{(I)}(\vec{r}, \vec{r}^{\prime}, \omega)$ be the electronic density-density
    response of object $\mathrm{O}_{I}$, with all Coulomb interactions within $\mathrm{O}
    _{I}$ included. In the absence of the inter-object Coulomb interactions, the
    response is a sum $\chi=\sum_{I=1}^{N}\chi^{(I)}$. In the presence of the
    inter-object Coulomb interaction $w_{IJ}$, we will assume that each object responds
    linearly to the potential generated by the other objects. Thus the dynamic
    electron density perturbation on $\mathrm{O}_{I}$ is
    $n^{(I)}=\chi^{(I)}\sum_{J}w_{IJ}\ n^{(J)}$, where products are spatial convolutions.
    The overall density response is then $\Tilde{\chi}=(1-\chi w)^{-1}\chi$ with
    inter-object interactions included and the inverse is taken with respect to
    convolution. By adiabatically switching on the interaction $w$ and using Feynman's
    theorem and the fluctuation-dissipation theorem, we obtain the inter-object
    free energy via the response functions at imaginary frequency, $\omega=i u$,
    \begin{equation}
        \begin{aligned}
            E & = K \sum_{u, r}\ln (1-\chi(i u) w)_{\vec{r}, \vec{r}}                                                                \\
              & =K \sum_{u, r}\ln \left(1-\overleftrightarrow{\alpha}(i u) \bullet \overleftrightarrow{T}\right)_{\vec{r}, \vec{r}}.
        \end{aligned}
    \end{equation}
    Here the logarithm and products (convolutions) are over the space of
    positions $\vec{r}$ (and summed over Cartesian indices $i,j = 1,2,3$ in the final
    expression containing $\overleftrightarrow{\alpha}\bullet \overleftrightarrow
    {T}$). $K$ is a positive constant. The imaginary frequency $u$ is summed
    over Matsubara frequencies or integrated over positive values, at finite or zero
    temperature respectively. The polarizability density
    $\overleftrightarrow{\alpha}$ is such that
    $\chi\left(\vec{r}, \vec{r}^{\prime}, \omega\right)=-|e|^{-2}\vec{\partial}_{r}
    . \vec{\partial}_{r^{\prime}}\overleftrightarrow{\alpha}(\vec{r}, \vec{r}^{\prime}
    , \omega)$
    and the Coulomb tensor is
    $T_{m n}=|e|^{2}\partial_{m}\partial_{n}^{\prime}\left|\vec{r}-\vec{r}^{\prime}
    \right|^{-1}$. This type of approach can be used to derive the RPA correlation
    energy\,\cite{angyan2020london}, the MBD vdW theory\,\cite{ambrosetti2014long},
    and the standard non-retarded Lifshitz theory\,\cite{Tim}. The operator
    logarithm can be Taylor expanded to give
    \begin{equation}
        E=-\sum_{n=2}^{\infty}K_{n}\mathrm{Tr}\left((\overleftrightarrow{\alpha}\bullet
        \overleftrightarrow{T})^{n}\right) \label{eqn:m9}
    \end{equation}
    where $K_{n}$ is a positive constant and
    $\operatorname{Tr}f \equiv \sum_{u, m}\int d\vec{r}\,f_{m m}(\vec{r}, \vec{r}
    , u)$. Noting that $\alpha=\sum_{I}\alpha^{(I)}$, we find Eq.\,(\ref{eqn:m9})
    contains $N$-object terms.
    The leading $N$-object term (i.e. that with the fewest Coulomb factors $T$) has
    $n=N$ and is of the form

    \begin{equation}
        \begin{aligned}
            E_{N}=-c_{N}\operatorname{Tr}\Bigg(\stackrel{\leftrightarrow}{\alpha}^{(1)}\bullet \stackrel{\leftrightarrow}{T}^{(1,2)}\bullet \stackrel{\leftrightarrow}{\alpha}^{(2)}\bullet \ldots \bullet \stackrel{\leftrightarrow}{T}^{(N-1, N)}\bullet \\
            \stackrel{\leftrightarrow}{\alpha}^{(N)}\bullet \stackrel{\leftrightarrow}{T}^{(N,1)}\Bigg)
        \end{aligned}
        \label{eqn:m10}
    \end{equation}
    where $c_{N}$ is positive constant, plus terms with the numbers $1,2, \ldots,
    N$ permuted but with none repeated. We will now establish the sign of the
    energy contribution in Eq.\,(\ref{eqn:m10}) for elongated uniaxially polarizable
    objects in the geometries shown in Figs.\,(\ref{fig:one}(a,b)).\\

    For long-wavelength excitations (corresponding to well-separated objects), the
    objects may be treated as translationally invariant in the $z$ direction (along
    the axis), with graininess (periodicity) in the $z$ direction acknowledged
    via electron effective masses $m^{*}$ from Bloch band theory. Then Eq.\,(\ref{eqn:m10})
    simplifies greatly in the space of wave numbers $q$, as follows. For objects
    polarizable only along the long $(z)$ axis as in Fig.\,(\ref{fig:one}) (the
    non-local polarizability can then be written as,
    \begin{equation}
        \begin{aligned}
            \stackrel{\leftrightarrow}{\alpha}^{(I)}\left(\vec{r}, \vec{r}^{\prime}, \omega\right)=(2 \pi)^{-1}\hat{z}\hat{z}\rho\left(\vec{r}_{\perp}\right) \rho\left(\vec{r}_{\perp}^{\prime}\right) \int_{0}^{\infty}dq \\
            \exp \left(i q\left(z-z^{\prime}\right)\right) \alpha_{\|}^{(I)}(q, \omega)
        \end{aligned}
    \end{equation}
    The function $\rho$ is the square of the transverse electronic wavefunction for
    the case of atomically-thin objects such as small-radius nanotubes or DNA,
    where electron motion is quantally confirmed in the $x$ and $y$ directions. For
    wider cylinders $\rho$ confines $\vec{r}_{\perp}$ to lie within the cylinder
    radius, and the present theory assumes the transverse electronic
    polarizability is negligible beside the longitudinal polarizability. For object
    separations much greater than the radius, we may take
    $\rho\left(\vec{r}_{\perp}\right)=\delta (x) \delta(y) = \delta(\vec{r}_{\perp}
    )$. The only property of $\alpha_{\|}^{(I)}$ required here is positivity,
    $\alpha_{\|}^{(I)}>0$. This is true for the standard low-$q$, low-$u$ model
    of longitudinally polarizable linear objects\,\cite{dobson2006asymptotics}: see
    Appendix B.\\

    For the geometry of Fig.\,(\ref{fig:one}a), the Fourier-transformed inter-object
    Coulomb tensor for two objects separated by distance $D$ is $\overleftrightarrow
    {T}(q)=\hat{z}\hat{z}T_{\|}(q)$ where $T_{\|}(q)=-|e|^{2}q^{2}K_{0}(q D)<0$.
    This is negative: a right-directed dipole on one object produces a left-directed
    field on a nearby parallel object, causing a contrary polarization of the
    second object corresponding to screening as indicated in Fig.\,(\ref{fig:one}a).
    For transversely polarizable linear structures in the geometry of Fig.\,(\ref{fig:one}b),
    we take
    $\overleftrightarrow{\alpha}^{(I)}(\vec{r}_{\perp}, \vec{r}_{\perp}{ }^{\prime}
    , q, i u)=\hat{x}\hat{x}\alpha_{\perp}{ }^{(I)}(i u) \delta(\vec{r}_{\perp})\delta
    (\vec{r}_{\perp}{ }^{\prime})$
    where $\alpha_{\perp}{ }^{(I)}>0$ and the $x$ axis points between the
    parallel objects as in Fig.\,(\ref{fig:one}b). For this case the relevant Fourier
    transformed Coulomb tensor is $\overleftrightarrow{T}(q)=\hat{x}\hat{x}T_{\perp}
    (q)$ where $T_{\perp}(q)=|e|^{2}d^{2}K_{0}(q D) / d D^{2}>0$. This is
    positive: an upward dipole on the lower object in Fig.\,(\ref{fig:one}b) produces
    an upward field on the upper object, causing the anti-screening dipole shown
    faint in the figure.

    The above models greatly simplify the calculation of the $N$-object term in
    Eq.\,(\ref{eqn:m10}). The spatial convolutions now become simple products in
    $q$ space, and the tensor products $\overleftrightarrow{\alpha}\bullet \overleftrightarrow
    {T}$ become simple products $\alpha_{\|}(q, i u) T_{\|}(q)$ or $\alpha_{\perp}
    (i u) T_{\perp}(q)$ for Fig.\,(\ref{fig:one}(a,b)) respectively. Knowing the
    signs of $\alpha$ and $T$ then allows determination of the sign of the $N$-object
    term of Eq.\,(\ref{eqn:m10}), as follows. For the geometry of Fig.\,(\ref{fig:one}a),
    we have $\alpha_{I}= \alpha_{\|}{ }^{(I)}>0, T^{(I,J)}=T_{\|}^{(I,J)}(q)<0$.
    The sign of of the $N$-object energy contribution in Eq.\,(\ref{eqn:m10}) is
    $\operatorname{sgn}\left(E_{N}\right) =-(+1)^{N}(-1)^{N}=-(-1)^{N}$. Thus the
    leading irreducible $N$-object contribution to the dispersion interaction for
    $N$ parallel linear, longitudinally polarizable objects is negative (attractive)
    for even $N$, and positive (repulsive) for odd $N$.

\begin{figure}[h]
        \centering
        \includegraphics[width=0.7\linewidth]{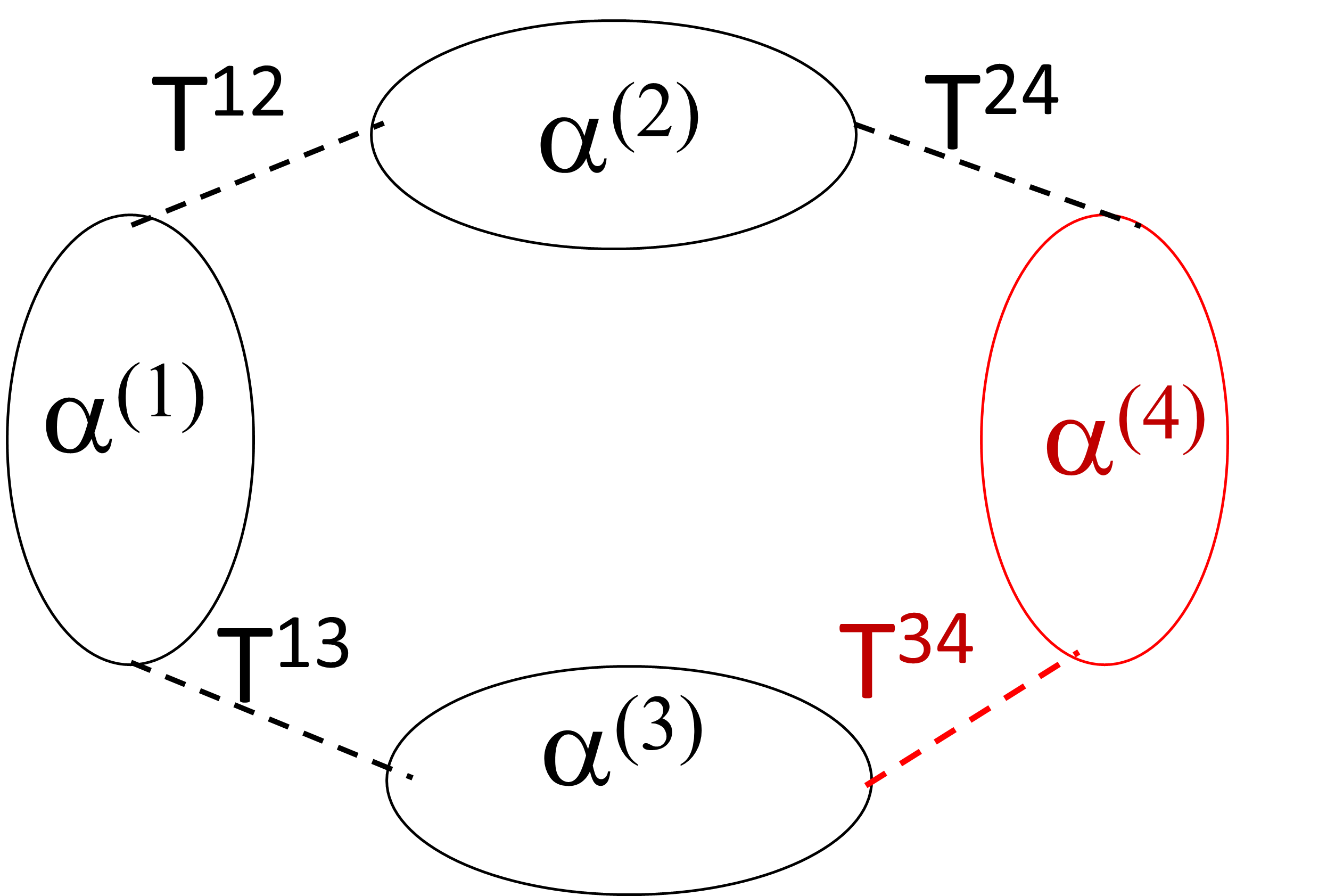}
        \caption{(Colors online) Feynman diagram for the leading irreducible 4-object
        vdW interaction Eq.\,(\ref{eqn:m10}). The red lines represent the
        screening/anti-screening of the interaction between O2 and O3 due to the
        introduction of the fourth-polarizable object O4, as per the qualitative
        argument of the previous section}
        \label{fig:two}
    \end{figure}

    By contrast, for the geometry of Fig.\,(\ref{fig:one}b), we have $\alpha_{I}=
    \alpha_{\perp}{ }^{(I)}>0, T^{(I,J)}=T_{\perp}{ }^{(I, J)}(q)>0$ and so Eq.\,(\ref{eqn:m10})
    is negative definite. Thus the irreducible $N$-object contribution to the dispersion
    interaction among $N$ parallel linear, objects that are polarizable in the
    $x$ direction of Fig.\,(\ref{fig:one}b) is negative (attractive) for all $N$.
    \\ For objects that are significantly polarizable in more than one direction,
    the above screening and anti-screening effects can compete, and no general prediction
    can be made for the sign of the $N$-object energy term. \\ The $N = 4$ case
    of Eq.\,(\ref{eqn:m10}) is illustrated in Fig.\,(\ref{fig:two}) by a Feynman-style
    diagram.

    \section{Confirmation from a plasma cylinder model}
    \label{sec4} 
\begin{figure}[!h]
        \centering
        \includegraphics[width=0.7\linewidth]{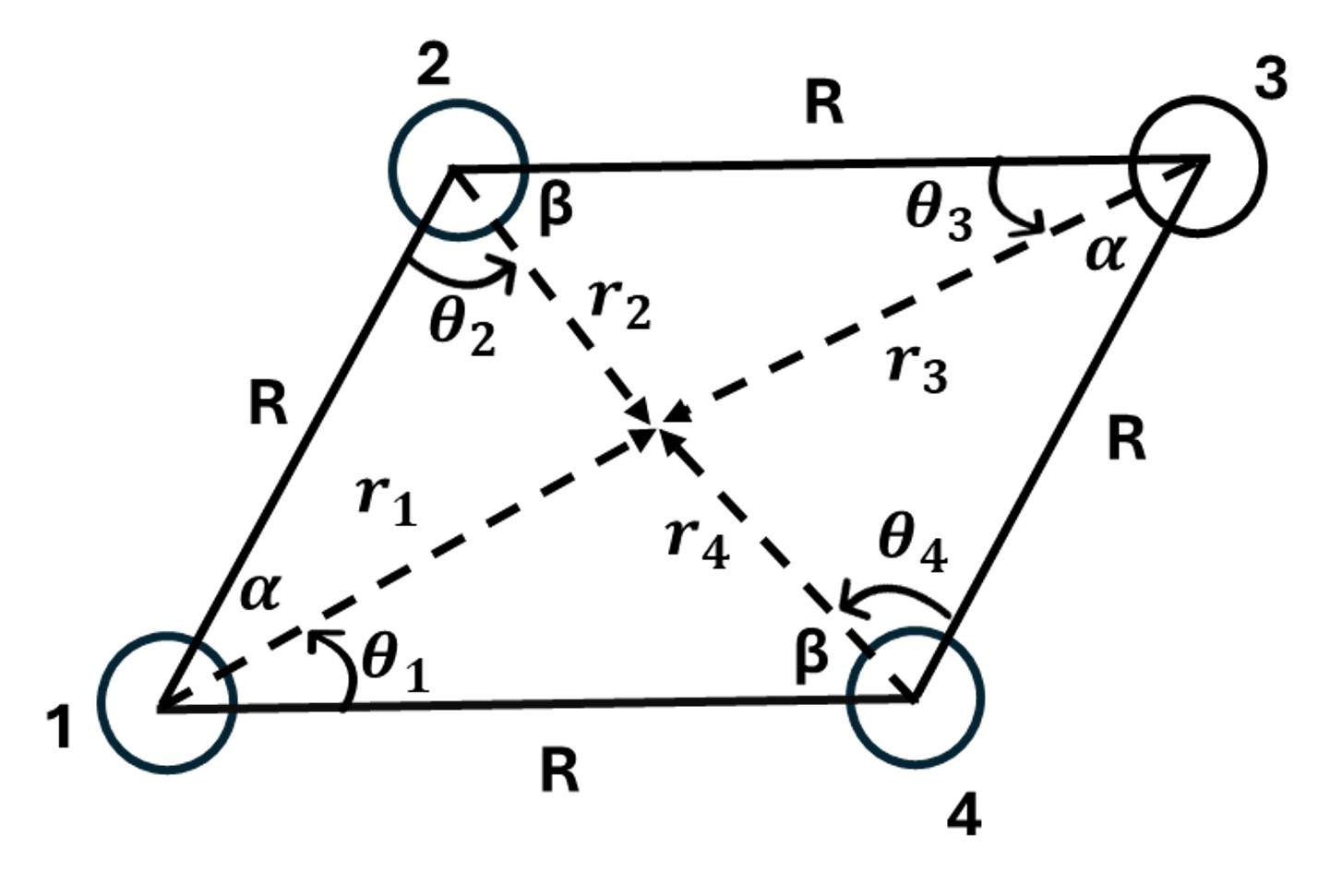}
        \caption{Schematic representation of four thin conducting cylinders at
        the vertices of a Rhombus within a vacuum chamber. $\alpha$ and $\beta$ are
        opposite angles. $d_{1}$ and $d_{2}$ are the diagonals of Rhombus, the distance
        between cylinder 1, cylinder 3 and cylinder 2 and cylinder 4.}
        \label{fig:three}
    \end{figure}

The last section analyzed the sign of the $N$-object dispersion
    energy term of elongated objects within a model that was rather general
    except that it was quasi-1-dimensional. We now confirm these results for $N =
    2,3,4$ within a specific model that is truly three-dimensional. This model
    is in fact where we first observed the alternating sign effect. We consider four
    identical parallel conducting cylinders (many atoms thick to avoid quantum
    effects discussed in the Appendix, with radius $a$ and length $L$ placed at the
    vertices of a Rhombus within a vacuum chamber as shown in Fig.\,(\ref{fig:three}).
    The separation between two consecutive cylinders is taken to be $R$. The
    long axes of all cylinders are aligned in the $z$ direction. In previous studies,
    for systems with two\,\cite{SubhojitPCCP2024} and three cylinders\,\cite{richmond1972many},
    a conduction process was described through a simple linearized hydrodynamic (electron
    plasma) model neglecting collisions between the charge carriers and incorporated
    a continuum method to characterize the dynamics of free charge carriers.
    This model could describe lightly doped semiconductors, for example. The
    electric field is obtained from an isotropic scalar potential $\Phi$, which satisfies
    Poisson's equation inside each cylinder and Laplace's equations elsewhere. (Note:
    the charge fluctuations are only allowed inside the cylinders.)

    The normalized solution for the potential inside the cylinders was derived
    by Davies {\it et al.}\,\cite{Davies1973} in terms of radial polar
    coordinates ($r_{i}, \theta_{i}$) and centered on the axis of cylinder $i$,

    \begin{equation}
        \begin{aligned}
            \Phi^{(i)}_{\text{in}}= \sum_{m}A^{(i)}_{m}\exp(i m\theta_{i}) \Big[ I_{m}(kr_{i}) - \gamma_{m}I_{m}(ur_{i})\Big] \\
            \exp[i(k z - \omega t)], \quad (i= 1,2,3,4)
        \end{aligned}
    \end{equation}
    where
    $\gamma_{m}= \frac{k \omega^{2}_{p}I^{\prime}(kb)}{u\omega^{2}I^{\prime}(ub)}$
    and $u^{2}= k^{2}+ (\omega^{2}_{p}- \omega^{2})/s^{2}$,
    and $\omega_{p}$ be plasma frequency, has denoted by
    $\omega_{p}^{2}= 4\pi n_{0}e^{2}/m$ and $s$ is isothermal sound velocity of
    the charge carriers, $s^{2}= m^{-1}(\partial p/\partial n)$, $n_{0}$ is the
    equilibrium density of free charge carriers, mass $m$ and $p$ be the pressure.
    This approach of Davies {\it et al.}\,\cite{Davies1973} means that the sound
    velocity appears in the dispersion relations in the cylinders, the propagation
    velocity of compressional waves playing an important role. Outside the
    cylinders, the fields are given as,
    \begin{equation}
        \Phi_{\text{ext}}= \sum_{i,m}B^{(i)}_{m}\exp(i m\theta_{i}) K_{m}(kr_{i})
        \exp[i(k z - \omega t)]
    \end{equation}
    where $I_{m}$ and $K_{m}$ are modified Bessel functions of first and second
    kind respectively in standard notations, and $A_{m}$'s, $B_{m}$'s are
    coefficients which we need to determine. The systematic procedure to connect
    these coefficients is to first represent the external potential in terms of one
    cylinder coordinates and then satisfy the necessary boundary conditions at surface
    of the cylinders discussed in the Appendix by stating that free charges don't
    assemble on the surface.
    Using Graf's addition theorem\,\cite{watson1958} for modified Bessel
    functions, we can express the potential outside all cylinders in the coordinates
    of cylinder 1 as,
    \begin{equation}
        \begin{aligned}
            \Phi^{(1)}_{\text{ext}}= \sum_{m}\Big[ B^{(1)}_{m}K_{m}(kr_{1}) + \sum_{m^{\prime}}\Big( B^{(2)}_{m^{\prime}}K_{m^{\prime}-m}(kR) e^{i m \alpha}+ B^{(4)}_{m^{\prime}} \\
            K_{m^{\prime}+ m}(kR) e^{i m \alpha}+ B^{(3)}_{m^{\prime}}K_{m^{\prime}-m}(kd_{1}) e^{i m \beta}\Big) I_{m}(kr_{1})                                                    \\
            e^{i m \theta_1}\exp[i(kz- \omega t)] \Big]
        \end{aligned}
    \end{equation}

    We derive in the Appendix coefficients using the boundary conditions including
    the continuity of $\Phi^{(1)}$ and $\partial \Phi^{(1)}/\partial n$ at the surface
    of cylinder 1 $(r_{1}= a)$. The same method is generalized including all
    cylinders. The theory is mathematically challenging but novel results are obtained
    in the ``thin'' cylinder approximation $R \gg a$. In this limit, only zeroth-order
    terms for small arguments in the expansion contribute. Hence the simplified
    dispersion relation ($\mathscr{D}(\omega) = 0$) for all surface modes can be
    analyzed in the Appendix C,
    \begin{equation}
        \begin{aligned}
            \mathscr{D}(\omega) = 1 - \underbrace{4 A^2 K^2_{0}(kR) - A^2 K^{2}_{0}(k d_1) - A^2 K^2_{0}(kd_2)}_{C^{(2)}} \\
            - \overbrace{4 A^3 K^2_{0}(kR) (K_{0}(kd_1) + K_{0}(kd_2))}^{C^{(3)}}                                         \\
            - \underbrace{3 A^4 K^2_{0}(kR) K_{0}(k d_1)K_{0}(k d_2) + A^4 K^{2}_{0}(k d_1)K^{2}_{0}(k d_2)}_{C^{(4)}}
        \end{aligned}
        \label{EqDispersionRelation}
    \end{equation}

    where
    \begin{equation}
        \begin{aligned}
            A = \frac{1}{2}(k\,a)^{2}\frac{\omega^2_{p}}{\omega^2 - k^2 s^2 \Big[1- \frac{1}{2} (a/\lambda_{D})^2 \ln(ka)\Big]}, \\
            ka \ll 1; \quad \lambda_{D}= \frac{s}{\omega_{p}}
        \end{aligned}
        \label{eqn:aaa}
    \end{equation}
    where $\lambda_{D}$ is Debye screening length. Formally the ground state interaction
    per unit length (for a cylinder with length $L$) can be written as\,\cite{Davies1973,Maha},
    \begin{equation}
        \begin{split}
            F(a,R,T) \simeq \frac{k_{B}T}{\pi}\sum_{n=0}^{\infty}{}' \int_{0}^{\infty}
            dk\,\ln \mathscr{D}(i \xi_{n})
        \end{split}
        \label{eqn:m299}
    \end{equation}
    where the prime indicates that the zero frequency term carries a weight 1/2
    and Matsubara frequency is $\xi_{n}=2 \pi k T n/\hbar$. In the large separation
    limit, the zero frequency term is the only surviving contribution and leads
    as we will demonstrate to entropic (classic) asymptotes that are attractive
    for 2 and 4-object interactions and repulsive for 3-object interactions. The
    low and high-temperature limits can be treated consistently by replacing the
    finite temperature free energy Matsubara frequency summation with a zero temperature
    frequency integration\,\cite{Dzya}. The high temperature-long distance
    limits are obtained by taking the zero frequency term in the Matsubara
    summation since in a retarded theory all finite frequency terms are then
    screened out by the finite velocity of light. We can derive the many-object terms
    by considering the relevant limits. We substitute Eq.\,(\ref{eqn:aaa}) into Eq.\,(\ref{eqn:m299})
    and expand the logarithm as $\ln(1-x) \simeq -x$ when $x \ll 1$; where $x$ can
    be assumed as a function the relevant Bessel functions ($K_{0}(kR)$ and
    $K_{0}(kR) \ll K_{0}(ka)$). We find that the many-object dispersion
    interaction energy per unit length at low temperatures is given by,
    \begin{equation}
        \begin{aligned}
            F(a,R) \simeq - \frac{\hbar}{2 \pi^2}\int_{0}^{\infty}d\xi \int_{0}^{\infty}dk [C^{(2)}+ C^{(3)}+ C^{(4)}]
        \end{aligned}
    \end{equation}
    the first term inside the integral in the last expression denotes two-object
    contribution where as the second and the third term are the three-object and
    four-object contributions to the total energy. Notably, we explore the 2, 3,
    and 4-objects interactions in the long-range non-retarded limits and in the corresponding
    long-range entropic limits. The later is valid at large separations and/or
    high temperatures.

    \begin{table*}
        [t]
        \centering
        \begin{tabular}{ p{3cm} p{3cm} p{5.25cm} p{5cm} }
            \hline
            \hline
            \multirow{2}{*}{\centering \quad System}        & \hspace{-0.2in}\multirow{2}{*}{\makecell{Approximations\\ (NR limit)}} & \multicolumn{2}{c}{Power-laws}                                                                                                \\
            \cline{3-4}                                     &                                                                        & High T                                                                                                                       & Low T                                                                                                            \\
            \hline
            \makecell{2 $||^{n}$ cylinders\\ contribution}  & \quad $a \ll \lambda_{D}$                                              & $-R^{-1}$                                                                                                                    & $-R^{-2}$                                                                                                        \\
            \hline
                                                            & \quad $a \gg \lambda_{D}$                                              & $-R^{-1}[\ln(R/a)]^{-2}$                                                                                                     & $-R^{-2}[\ln(R/a)]^{-3/2}$                                                                                       \\
            \hline
            \makecell{3 $||^{n}$ cylinders\\ contributions} & \quad $a \ll \lambda_{D}$                                              & \quad $R^{-1}$                                                                                                               & \quad $R^{-2}$                                                                                                   \\
            \hline
                                                            & \quad $a \gg \lambda_{D}$                                              & \quad $R^{-1}[\ln(R/a)]^{-3}$                                                                                                & \quad $R^{-2}[\ln(R/a)]^{-5/2}$                                                                                  \\
            \hline
            \makecell{4 $||^{n}$ cylinders\\ contributions} & \quad $a \ll \lambda_{D}$                                              & \makecell{$-R^{-1}$\\$\Big[3\, g^{\prime}(\alpha, \beta, R) - X^{\prime}(\alpha, \beta, R)\Big]$}                            & \makecell{$-R^{-2}$\\$\Big[3\, g(\alpha, \beta, R) - X(\alpha, \beta, R)\Big]$}                                  \\
            \hline
                                                            & \quad $a \gg \lambda_{D}$                                              & \makecell{$-R^{-1}[\ln(R/a)]^{-4}$\\$\Big[3\, g^{\prime\prime}(\alpha, \beta, R) - X^{\prime\prime}(\alpha, \beta, R)\Big]$} & \makecell{$-R^{-2}[\ln(R/a)]^{-7/2}$\\$\Big[3\, \Tilde{g}(\alpha, \beta, R) - \Tilde{X}(\alpha, \beta, R)\Big]$} \\
            \hline
            \hline
        \end{tabular}\\
        \caption{\label{SubhojitPowerLawTable} Asymptotic power-law dependency
        for van der Waals interaction for different cylindrical configurations
        for four-object interaction. The description and detailed derivation of all
        these terms are given in the Appendix. NR signifies non-retarded limit.}
    \end{table*}

    The asymptotic limits of the multi-objects interactions involved in the case
    of 4 thin conducting cylinders are presented in Table.\,\ref{SubhojitPowerLawTable}.
    The 2-object contribution is attractive as described in detail multiple times\,\cite{Davies1973,SubhojitPCCP2024,Richmond1972}.
    In contrast, the case of the repulsive 3-object term is much less well described.
    Only a few final asymptotic results are given by Richmond and Davies\,\cite{richmond1972many}.
    in the Appendix, we present the derivations including enough technical
    details to obtain the 3-object results. We demonstrate for the first time that
    the 4-object force is attractive. Notably, we explore the results up to 4-object
    interactions valid both in the non-retarded limit and in the long-range zero
    frequency (entropic) limit. The signs of the 2, 3 and 4 body energies from Table.\,\ref{SubhojitPowerLawTable}
    are negative, positive and negative respectively, confirming, within a our 3d
    model, the alternating signs predicted in earlier Sections based on a 1D
    model.

    \section{Summary}
    Filamentary structures are ubiquitous in nano- and bio-science, and many examples have a high conductivity. These systems interact via dispersion forces, causing then to bundle into parallel arrays. The highly polarizable, anisotropic nature of conducting filaments means that the beyond-pairwise terms are large in a perturbative dispersion energy calculation for arrays of filaments.

    We have studied these beyond-pairwise interactions. The new results from our work are as follows. (a) The beyond-pairwise contributions to the many-filament dispersion interaction alternate in sign, with the leading $N$-filament contribution having sign $(-1)^{N+1}$. That is, odd-$N$ contributions are repulsive and even-$N$ contributions are attractive. These results were proven within a quasi-one-dimensional model of a filament. This simple result for filamentary systems contrasts with the case of general geometry, where detailed calculations are required to determine these signs. (b) We have provided a qualitative analysis, based on the concepts of Coulomb screening and anti-screening, allowing these trends to be understood intuitively. (c) We have given the first analysis of a four-filament system beyond the quasi-one-dimensional approximation, verifying the above-mentioned sign alternation within a three-dimensional plasma cylinder model. (d) We have given the power-law decay falloff of $N^{\rm{th}}$ dispersion energy term, for $N$=2,3 and 4. (e) The alternating signs make it difficult to ascertain convergence in a perturbative numerical analysis of the multi-filaments dispersion interaction. Therefore a non-perturbative approach will be needed for quantitative analysis.

    \section{Implications for future work}
    \label{BiologyNano}

     We are currently investigating non-perturbative approaches for the dispersion energy of multi-filament bundles. Semi-analytic results appear to be feasible for symmetric arrays within a quasi-one-dimensional model.

    The present analysis used continuum models, meaning that the results are reliable when the objects are separated by more than a few atomic lattice spacings. The well-known MBD approach\,\cite{ambrosetti2014long}
    provides efficient non-perturbative numerical vdW energy modeling at the lumped atomic level, without the need for high-level quantum chemical methods. MBD correctly describes the effects of discrete atomic structure on the vdW interaction near to contact. It also captures the more distant vdW interaction in most cases. Unfortunately, however, MBD doesn't account for type-C vdW non-additivity\,\cite{dobson2014}
    and so misses the anomalous long-ranged vdW interactions of conducting filaments\,\cite{ambrosetti2016wavelike}.

   A new approach, MBD+C\,\cite{dobson2023mbd+}, accounts seamlessly for both regimes including the case of conducting low-dimensional structures such as filaments. MBD+C is still under development but promises efficient and reliable numerical modeling of vdW effects in the ubiquitous filamentary structures considered here.

    \appendix

    \section{{Brief discussion on quantum effects}}

    {Our work in general is relevant to both quantum and classical many-body interactions between elongated particles. For those systems where we use a plasma model we}
    primarily consider cylinders many atoms thick so we can avoid quantum
    effects and assume electron densities corresponding to semiconductor
    cylinders lightly doped. The electron clouds can then be treated as classical
    plasma where electrons can move freely within the cylindrical barriers\,\cite{Richmond1972,Davies1973}.
    Electron degeneracy in densely packed biological systems occurs when the
    quantum states fill up to a large fraction of the Fermi level. They do then obey
    Fermi-Dirac rather than a Maxwell-Boltzmann distribution\,\cite{AshcroftMermin,Mahan}.
    Even in lightly doped semiconductors, at sufficiently low temperatures,
    electrons can become degenerate. To treat the system classically, the
    temperature must be high enough, and the electron density low enough, to
    ensure that quantum effects are negligible\,\cite{AshcroftMermin,Mahan}. However,
    as has been seen in the past, e.g. for van der Waals interaction between a
    pair of two-dimensional electron gas systems, quantum effects sometimes have
    less impact on the long-range vdW asymptotes than expected\,\cite{BostromSerneliusPhysRevB.61.2204}.

    \section{Electronic response of 1D electrons}
    A rather general model for the response of a quasi-1D linear objects\,\cite{dobson2006asymptotics},
    \begin{equation}
        \begin{aligned}
            \alpha_{\|}^{(I)}=-|e|^{2}q^{-2}\chi^{(I)}(q, \omega=i u)=+|e|^{2}n_{0}\left(m^{*}\right)^{-1} \\
            \left(u^{2}+\omega_{1 D}^{2}(q)+\omega_{o}^{2}\right)^{-1}
        \end{aligned}
        \label{jfd}
    \end{equation}
    Here $n_{0}$ is the number of polarizable electrons per unit length of
    object $O_{I}$, $\omega_{0}$ is the band-gap frequency which vanishes for metals,
    and will be assumed small here, giving a large parallel polarizability at small
    $q$ and $u$. $\omega_{1 D}(q)$ is the one-dimensional plasma frequency which
    $\rightarrow 0$ as $q \rightarrow 0$\,\cite{dobson2006asymptotics}. Eq.\,(\ref{jfd})
    can be obtained from the hydrodynamic arguments, also from the long-wavelength
    limit of microscopic Bloch electron response theory.

    \section{Derivation of dispersion relation for N=4 objects for 3D plasma
    cylinder model}
    \begin{figure}[!h]
        \centering
        \includegraphics[width=0.7\columnwidth]{SM1.pdf}
        \caption{ (Colors online) Schematic representation of four parallel cylinders
        where 1, 2, 3 and 4 denote cylinder numbers put at the vertices of a Rhombus
        and $\alpha$ and $\beta$ are opposite angles. $d_{1}$ and $d_{2}$ are
        the diagonals of Rhombus, the distance between cylinder 1, cylinder 3
        and cylinder 2 and cylinder 4.}
        \label{fig:scheme1}
    \end{figure}
    The normalized solution for the potential inside the cylinders was derived by
    Davies {\it et al.} in terms of radical polar coordinates ($r_{i}, \theta_{i}$)
    and centered on the axis of cylinder $i$,
    \begin{equation}
        \begin{aligned}
            \Phi^{(i)}_{\text{in}}= \sum_{m}A^{(i)}_{m}\exp(i m\theta_{i}) \Big[ I_{m}(kr_{i}) - \gamma_{m}I_{m}(ur_{i})\Big] \\
            \exp[i(k z - \omega t)], \quad (i= 1,2,3,4)
        \end{aligned}
        \label{eqn:one}
    \end{equation}
    where $\gamma_{m}= \frac{k A^{2}_{p}I^{\prime}(kb)}{uA^{2}I^{\prime}(ub)}$
    and $u^{2}= k^{2}+ (A^{2}_{p}- A^{2})/s^{2}$. Outside the cylinders, the
    fields are given as,
    \begin{equation}
        \Phi_{\text{ext}}= \sum_{i,m}B^{(i)}_{m}\exp(i m\theta_{i}) K_{m}(kr_{i})
        \exp[i(k z - \omega t)]
    \end{equation}
    where $I_{m}$ and $K_{m}$ are modified Bessel functions of first and second kind
    respectively in standard notations. To represent the external potential in
    terms of the coordinates of one cylinder, we can use Graf's summation formula
    for Bessel functions as, (here we have shown only transformation of cylinder
    2 coordinates in terms of coordinates of cylinder 1)
    \begin{equation}
        \begin{aligned}
            K_{m}(kr_{2}) \exp({im\theta_{2}}) = \sum_{m^{\prime} = -\infty}^{\infty}K_{m^{\prime} - m}(kR) I_{m}(kr_{1}) \\
            \exp({im\theta_{1}})                                                                                          
        \end{aligned}
    \end{equation}

    Now we can express the potential outside the cylinders in the coordinates of
    cylinder 1 as,

    \begin{equation}
        \begin{aligned}
            \Phi^{(1)}_{\text{ext}}= \sum_{m}\Big[ B^{(1)}_{m}K_{m}(kr_{1}) + \sum_{m^{\prime}}\Big( B^{(2)}_{m^{\prime}}K_{m^{\prime}-m}(kR) e^{i m \alpha}+ B^{(4)}_{m^{\prime}} \\
            K_{m^{\prime}+ m}(kR) e^{i m \alpha}+ B^{(3)}_{m^{\prime}}K_{m^{\prime}-m}(kd_{1}) e^{i m \beta}\Big) I_{m}(kr_{1})                                                    \\
            e^{i m \theta_1}\exp[i(kz- \omega t)] \Big]
        \end{aligned}
        \label{eqn:4}
    \end{equation}

    Similarly, the external field in terms of the coordinates of cylinder 2,
    cylinder 3 and cylinder 4 can be written as,
    \begin{equation}
        \begin{aligned}
      \Phi^{(2)}_{\text{ext}}= \sum_{m}\Big[ B^{(2)}_{m}K_{m}(kr_{2}) + \sum_{m^{\prime}}\Big( B^{(1)}_{m^{\prime}}K_{m^{\prime}+m}(kR) e^{i m \beta}+ B^{(3)}_{m^{\prime}} \\
            K_{m^{\prime}-m}(kR) e^{i m \beta}+ B^{(4)}_{m^{\prime}}K_{m^{\prime}-m}(kd_{2}) e^{i m \alpha}\Big) I_{m}(kr_{2})\\
            e^{i m \theta_2}\exp[i(kz- \omega t)] \Big] 
        \end{aligned}
    \end{equation}
    \begin{equation}
        \begin{aligned}
            \Phi^{(3)}_{\text{ext}}= \sum_{m}\Big[ B^{(3)}_{m}K_{m}(kr_{3}) + \sum_{m^{\prime}}\Big( B^{(4)}_{m^{\prime}}K_{m^{\prime}-m}(kR) e^{i m \alpha}+ B^{(2)}_{m^{\prime}} \\ K_{m^{\prime}+m}(kR) 
            e^{i m \alpha}+ B^{(1)}_{m^{\prime}}K_{m^{\prime}-m}(kd_{1}) e^{i m \beta}\Big) I_{m}(kr_{3})\\ e^{i m \theta_3}\exp[i(kz- \omega t)] \Big]
        \end{aligned}
    \end{equation}
    \begin{equation}
        \begin{aligned}
            \Phi^{(4)}_{\text{ext}}= \sum_{m}\Big[ B^{(4)}_{m}K_{m}(kr_{4}) + \sum_{m^{\prime}}\Big( B^{(1)}_{m^{\prime}}  K_{m^{\prime}-m}(kR) e^{i m \beta}+ B^{(2)}_{m^{\prime}} \\ K_{m^{\prime}-m}(kd_{2}) 
            e^{i m \alpha}+ B^{(3)}_{m^{\prime}}K_{m^{\prime} + m}(kR) e^{i m \beta}\Big) I_{m}(kr_{4}) \\ e^{i m \theta_4}\exp[i(kz- \omega t)] \Big]
        \end{aligned}
    \end{equation}

    Considering Eq.\,(\ref{eqn:one}) and Eq.\,(\ref{eqn:4}) for the first
    cylinder, and enforcing the potential continuity, namely
    $\Phi^{(1)}_{\text{in}}= \Phi^{(1)}_{\text{ext}}$ at $r_{1}= a$, we obtain
    \begin{equation}
        \begin{aligned}
            A^{(1)}_{m}\underbrace{\Big[ I_{m}(ka) - \gamma_{m}\, I_{m}(ua)\Big]}_{\tilde{X}}= B^{(1)}_{m}K_{m}(ka) + \sum_{m^{\prime}}\Big( B^{(2)}_{m^{\prime}}K_{m^{\prime} - m }(kR) \\ e^{im \alpha}+ 
            B^{(3)}_{m^{\prime}}K_{m^{\prime} - m }(kd_{1}) e^{im\beta}+ B^{(4)}_{m^{\prime}}K_{m^{\prime}+ m}(kR) e^{i m \alpha}\Big) I_{m}(ka)
        \end{aligned}
        \label{eqn:8}
    \end{equation}
    and the remaining boundary condition ensures the continuity of $\partial \Phi
    /\partial r_{1}$ at the boundary of cylinder 1 that gives
    \begin{equation}
        \begin{aligned}
           A^{(1)}_{m}\overbrace{\Big[ k I^{\prime}_{m}(ka) - \gamma_{m} u \, I^{\prime}_{m}(ua)\Big]}^{\tilde{Y}}= B^{(1)}_{m}k K^{\prime}_{m}(ka) + \sum_{m^{\prime}}k \Big( B^{(2)}_{m^{\prime}}K_{m^{\prime} - m }(kR) \\ e^{im \alpha} 
            + B^{(3)}_{m^{\prime}}K_{m^{\prime} - m }(kd_{1}) e^{im\beta}+ B^{(4)}_{m^{\prime}}K_{m^{\prime}+ m}(kR) e^{i m \alpha}\Big) I^{\prime}_{m}(ka)
        \end{aligned}
        \label{eqn:9}
    \end{equation}
    Eliminating $A^{(1)}_{m}$ from both the Eqs.\,(\ref{eqn:8}) and (\ref{eqn:9}),
    we obtain
    \begin{equation}
        \begin{aligned}
          B^{(1)}_{m}= \sum_{m^{\prime}}\underbrace{\Bigg[\frac{ k I^{\prime}_{m}(ka) \frac{\tilde{X}}{\tilde{Y}} - I_{m}(ka)}{K_{m}(ka) - k \frac{\tilde{X}}{\tilde{Y}} K^{\prime}_{m}(ka)}\Bigg]}_{A}\Big( B^{(2)}_{m^{\prime}}K_{m^{\prime} - m }(kR) e^{im \alpha} \\
           + B^{(3)}_{m^{\prime}}K_{m^{\prime} - m }(kd_{1}) e^{im\beta}+ B^{(4)}_{m^{\prime}}K_{m^{\prime}+ m}(kR) e^{i m \alpha}\Big)
        \end{aligned}
        \label{eqn:10}
    \end{equation}
    If we carry out the same procedure for cylinder 2, cylinder 3, and cylinder 4,
    we will be able to obtain a couple of expressions for the coefficients
    $B^{(2)}_{m}$, $B^{(3)}_{m}$ and $B^{(4)}_{m}$ in terms of $B^{(i)}_{m^{\prime}}$,
    $(i =1, 2, 3, 4)$ similar to Eq.\,(\ref{eqn:10}). These coefficients can be
    precisely represented in a matrix form as, $\widetilde{\gamma}= M \widetilde{\gamma}
    ^{\prime}$, where the matrix $M$ is given as,
    $M=$
    \begin{figure*}
        \begin{subequations}
            \begin{gather}
                \begin{pmatrix}
                    \begin{NiceArray}{:c:c:c:c:c:c}\\ \hdottedline\\ 0 & A e^{im\alpha} \sum_{m^{\prime}} K_{m^{\prime} - m }(kR) & A e^{im\beta} \sum_{m^{\prime}} K_{m^{\prime} - m }(kd_1) & A e^{im\alpha} \sum_{m^{\prime}} K_{m^{\prime}+ m }(kR) \\ \hdottedline\\ A e^{im\beta} \sum_{m^{\prime}} K_{m^{\prime} + m }(kR) & 0 & A e^{im\beta} \sum_{m^{\prime}} K_{m^{\prime} - m }(kR) & A e^{im\alpha} \sum_{m^{\prime}} K_{m^{\prime} - m }(kd_2) \\ \hdottedline\\ A e^{im\beta} \sum_{m^{\prime}} K_{m^{\prime} - m }(kd_1) & A e^{im\alpha} \sum_{m^{\prime}} K_{m^{\prime} + m }(kR) & 0 & A e^{im\alpha} \sum_{m^{\prime}} K_{m^{\prime} - m }(kR) \\ \hdottedline\\ A e^{im\beta} \sum_{m^{\prime}} K_{m^{\prime} - m }(kR) & A e^{im\alpha} \sum_{m^{\prime}} K_{m^{\prime} - m }(kd_2) & A e^{im\beta} \sum_{m^{\prime}} K_{m^{\prime} + m }(kR) & 0 \\ \hdottedline\\\end{NiceArray}
                \end{pmatrix}
                \label{eqn:15a}\\ \widetilde{\gamma}=
                \begin{pmatrix}
                    \vdots      \\
                    B^{(1)}_{m} \\
                    B^{(2)}_{m} \\
                    B^{(3)}_{m} \\
                    B^{(4)}_{m} \\
                    \vdots
                \end{pmatrix}
                \quad \quad \& \quad \quad \widetilde{\gamma}^{\prime}=
                \begin{pmatrix}
                    \vdots               \\
                    B^{(1)}_{m^{\prime}} \\
                    B^{(2)}_{m^{\prime}} \\
                    B^{(3)}_{m^{\prime}} \\
                    B^{(3)}_{m^{\prime}} \\
                    \vdots
                \end{pmatrix}\label{eqn:15b}
            \end{gather}
        \end{subequations}
    \end{figure*}
    We now derive an exact dispersion relation using the scattering matrix $\bf{M}$
    referenced in Eq.\,(\ref{eqn:15a}), which establishes the surface modes as follows,
    \begin{equation}
        \mathscr{D}(\omega) \equiv \Det( \I - M) = 0 \label{eqn:16}
    \end{equation}
    We are only interested in ``thin cylinder'' approximation and the ground
    state interaction because with increasing $m, m^{\prime}$, the matrix elements
    decrease rapidly. Now if we evaluate this determinant, we see a compact and simplified
    expression of this dispersion relation which is

    \begin{equation}
        \begin{aligned}
            \mathscr{D}(\omega) = 1 - 4 A^{2}K^{2}_{0}(kR) - A^{2}K^{2}_{0}(k d_{1}) - A^{2}K^{2}_{0}(kd_{2})- \\
            4 A^{3}K^{2}_{0}(kR) (K_{0}(kd_{1})+ K_{0}(kd_{2})) -                                              \\
            3 A^{4}K^{2}_{0}(kR) K_{0}(k d_{1})K_{0}(k d_{2}) + A^{4}K^{2}_{0}(k d_{1})K^{2}_{0}(k d_{2})
        \end{aligned}
    \end{equation}
    where
    \begin{equation}
        A = \frac{1}{2}(k\,a)^{2}\frac{\omega^{2}_{p}}{\omega^{2}- k^{2}s^{2}\Big[1-
        \frac{1}{2}(a/\lambda_{D})^{2}\ln(ka)\Big]}, \quad ka \ll 1; \quad \lambda
        _{D}= \frac{s}{\omega_{p}}
    \end{equation}
    Formally the ground state interaction per unit length (for a cylinder with length
    $L$) can be written as\,\cite{Davies1973,Maha},
    \begin{equation}
        \begin{split}
            F(a,R,T) \simeq \frac{k_{B}T}{\pi}\sum_{n=0}^{\infty}{}' \int_{0}^{\infty}
            dk\,\ln \mathscr{D}(i \xi_{n})
        \end{split}
        \label{eqn:299}
    \end{equation}
    where the Matsubara frequency $\xi_{n}=2 \pi k T n/\hbar$. The extra factor of 1/2 on the $n=0$ Matsubara contribution is well known and is often represented by putting a prime on the Matsubara summation. $\mathscr{D}$ contains the boundary-condition specific scattering information. This formalism is standard in the finite-temperature quantum field theory treatements of the Casimir effect.

    \section{Derivation of diagonal elements $d_{1}$ and $d_{2}$ in terms of $R$,
    $\alpha$, and $\beta$}
    There are two ways to calculate the diagonals of Rhombus. Here we are going to
    list both of them in a simple manner.
    \begin{enumerate}
        \item \underline{Laws of Sine} : The formula for laws of Sine is written
            as
            \begin{equation}
                \frac{R}{\sin(\alpha/2)}= \frac{d_{1}}{\sin(\beta)}= \frac{R}{\sin(\alpha/2)}
            \end{equation}
            \begin{figure}[!h]
                \centering
                \includegraphics[width=0.4\columnwidth]{triangle.pdf}
                \caption{ (Colors online) Schematic figure for determining the diagonal
                elements.}
                \label{fig:triangle}
            \end{figure}
            we know that $\alpha + \beta = 180^{\circ}$. Then
            \begin{equation}
                \frac{R}{\sin(\alpha/2)}= \frac{d_{1}}{2 \sin(\alpha/2)\cos(\alpha/2)}
                \implies \boxed{d_{1} = 2 R \cos(\alpha/2)}
            \end{equation}
            Similarly other diagonal $d_{2}$ is $\boxed{d_2 = 2 R \cos(\beta/2)}$.

        \item Using triangle formula
            \begin{equation}
                \begin{aligned}
                    d_{1} & = \sqrt{2R^2- 2R^2 \cos(\beta)}                         \\
                          & = R\sqrt{2(1- \cos(\beta))}                             \\
                          & = 2 R \cos(\alpha/2), \quad \beta = 180^{\circ}- \alpha
                \end{aligned}
            \end{equation}
    \end{enumerate}

    \section{Two-object contribution}

    Now we will focus on the two-object energy contribution for the zero
    temperature limit, which can be defined as $F^{(2)}$ [using
    $\ln(1-x) \simeq - x$)],
    \begin{equation}
        \begin{aligned}
            F^{(2)} & \simeq - \frac{\hbar}{2 \pi^2}\int_{0}^{\infty}d\xi \int_{0}^{\infty}dk \Big[4 A^{2}K^{2}_{0}(kR) + A^{2}K^{2}_{0}(k d_{1}) + A^{2}K^{2}_{0}(kd_{2})\Big]
        \end{aligned}
    \end{equation}
    Calculating the frequency integration using Mathematica software, we
    obtained
    \begin{equation}
        \begin{aligned}
            F^{(2)} & \simeq - \frac{\hbar \omega_{p} a^4}{32 \pi \lambda^{3}_{D}}\Bigg(\int_{0}^{\infty}dk k \, \frac{4 K^{2}_{0}(kR) + K^{2}_{0}(kd_1) +K^{2}_{0}(kd_2)}{\Big[1- \frac{1}{2} (a/\lambda_{D})^2 \ln(ka)\Big]^{3/2}}\Bigg)
        \end{aligned}
        \label{eqn:20}
    \end{equation}
    when $a \ll \lambda_{D}$, we can drop the denominator of Eq.\,(\ref{eqn:20})
    and therefore it yields
    \begin{equation}
        \begin{aligned}
            F^{(2)} & \simeq - \frac{\hbar \omega_{p} a^4}{64 \pi \lambda^{3}_{D}R^2}\Bigg[ 4 + \frac{\sec^2(\alpha/2)}{4}+ \frac{\sec^2(\beta/2)}{4}\Bigg]
        \end{aligned}
    \end{equation}
    and for $a \gg \lambda_{D}$, the maximum contribution in the integral Eq.\,(\ref{eqn:20})
    will come from the region where $k \lesssim R^{-1}$, hence using the approach
    described in Ref.\,\cite{SubhojitPCCP2024} we obtained
    \begin{equation}
        \begin{aligned}
            F^{(2)} & \simeq - \frac{\hbar \omega_{p} a}{8\sqrt{2} \pi R^2 [\ln(R/a)]^{3/2}}\Bigg[ 4 + \frac{\sec^2(\alpha/2)}{8}+ \frac{\sec^2(\beta/2)}{8}\Bigg]
        \end{aligned}
    \end{equation}
    \subsection{Zero-frequency contribution in the free energy}

    \begin{equation}
        \begin{aligned}
            F^{(2)}_{n=0} & \simeq - \frac{k_{B}T a^4}{8 \pi \lambda^{4}_{D}}\Bigg(\int_{0}^{\infty}dk \, \frac{4 K^{2}_{0}(kR) + K^{2}_{0}(kd_1) +K^{2}_{0}(kd_2)}{\Big[1- \frac{1}{2} (a/\lambda_{D})^2 \ln(ka)\Big]^{2}}\Bigg)
        \end{aligned}
    \end{equation}

    when $a \ll \lambda_{D}$
    \begin{equation}
        \begin{aligned}
            F^{(2)}_{n=0} & \simeq - \frac{\pi k_{B}T a^4}{32 \lambda^{4}_{D}R}\Bigg[ 4 + \frac{\sec(\alpha/2)}{2}+ \frac{\sec(\beta/2)}{2}\Bigg]
        \end{aligned}
    \end{equation}
    for $a \gg \lambda_{D}$
    \begin{equation}
        \begin{aligned}
            F^{(2)}_{n=0} & \simeq - \frac{\pi k_{B}T }{8 R\, [\ln(R/a)]^2}\Bigg[ 4 + \frac{\sec(\alpha/2)}{2}+ \frac{\sec(\beta/2)}{2}\Bigg]
        \end{aligned}
    \end{equation}
    \section{Three-object contribution}
    Three-object contribution can be compelled as $F^{(3)}$,
    \begin{equation}
        \begin{aligned}
            F^{(3)} & \simeq - \frac{\hbar}{2 \pi^2}\int_{0}^{\infty}d\xi \int_{0}^{\infty}dk \Big[4 A^{3}K^{2}_{0}(kR) (K_{0}(kd_{1}) + K_{0}(kd_{2}))\Big]
        \end{aligned}
    \end{equation}
    \begin{equation}
        \begin{aligned}
            F^{(3)} & \simeq \frac{3\hbar \omega_{p} a^6}{64 \pi \lambda^{5}_{D}}\Bigg(\int_{0}^{\infty}dk\, k \, \frac{ K^2_{0}(kR) (K_{0}(kd_1) + K_{0}(kd_2))}{\Big[1- \frac{1}{2} (a/\lambda_{D})^2 \ln(ka)\Big]^{5/2}}\Bigg)
        \end{aligned}
    \end{equation}
    When $a \ll \lambda_{D}$
    \begin{equation}
        \begin{aligned}
            F^{(3)} & \simeq \frac{3 \hbar \omega_{p} a^6}{64 \pi \lambda^5_{D} R^2}\, f(\alpha, \beta)
        \end{aligned}
    \end{equation}
    and for $a \gg \lambda_{D}$
    \begin{equation}
        \begin{aligned}
            F^{(3)} & \simeq \frac{3 \hbar \omega_{p} a}{64 \pi}\frac{4\sqrt{2}}{R^2[\ln(R/a)]^{\frac{5}{2}}}\, f(\alpha, \beta)
        \end{aligned}
    \end{equation}
    where
    \begin{equation}
        \begin{aligned}
            f(\alpha, \beta) = 0.22\, \Bigg[\MeijerG{3,2}{3,3}{1, 1, 1.5\\ 1,1,1}{\sec^2(\alpha/2)}+ \\
            \MeijerG{3,2}{3,3}{1, 1, 1.5\\ 1,1,1}{\sec^2(\beta/2)}\Bigg]
        \end{aligned}
    \end{equation}
    Here the $G$-function is Meijer's generalized $G$ function.
    \subsection{Zero-frequency contribution in the free energy}
    \begin{equation}
        \begin{aligned}
            F^{(3)}_{n=0} & \simeq \frac{k_{B}T a^6}{4 \pi \lambda^{6}_{D}}\Bigg(\int_{0}^{\infty}dk \, \frac{ K^2_{0}(kR) (K_{0}(kd_1) + K_{0}(kd_2))}{\Big[1- \frac{1}{2} (a/\lambda_{D})^2 \ln(ka)\Big]^{3}}\Bigg)
        \end{aligned}
    \end{equation}
    for $a\ll \lambda_{D}$
    \begin{equation}
        F^{(3)}_{n=0}\simeq \frac{k_{B}T a^{6}}{8 \lambda^{6}_{D}R}\, Y(\alpha, \beta
        )
    \end{equation}
    for $a\gg \lambda_{D}$
    \begin{equation}
        F^{(3)}_{n=0}\simeq \frac{k_{B}T}{R[\ln(R/a)]^{3}}\,Y(\alpha, \beta)
    \end{equation}
    where
    \begin{equation}
        \begin{aligned}
            Y(\alpha, \beta) = \Bigg[ K\Big(\frac{1}{2}(1-\sin(\alpha/2)\Big) K\Big(\frac{1}{2}(1+\sin(\alpha/2)\Big) + \\
            K\Big(\frac{1}{2}(1-\sin(\beta/2)\Big) K\Big(\frac{1}{2}(1+\sin(\beta/2)\Big) \Bigg]
        \end{aligned}
    \end{equation}
    $K$ is complete elliptic integral of first kind.
    \section{Four-object contribution}

    Four-object contribution can be compelled as $F^{(4)}$,
    \begin{equation}
        \begin{aligned}
           F^{(4)} \simeq - \frac{\hbar}{2 \pi^2}\int_{0}^{\infty}d\xi \int_{0}^{\infty}dk \Big[3 A^{4}K^{2}_{0}(kR) K_{0}(k d_{1})K_{0}(k d_{2}) \\
            - A^{4}K^{2}_{0}(k d_{1})K^{2}_{0}(k d_{2})\Big]
        \end{aligned}
    \end{equation}
    \begin{equation}
        \begin{aligned}
           F^{(4)}  \simeq -\frac{5\hbar \omega_{p} a^8}{1024 \pi \lambda^{7}_{D}}\Bigg(\int_{0}^{\infty}\frac{ dk\, k \, g(k)}{\Big[1- \frac{1}{2} (a/\lambda_{D})^2 \ln(ka)\Big]^{7/2}}\Bigg)
        \end{aligned}
    \end{equation}
    where
   $g(k) = \big(3 K^{2}_{0}(kR) K_{0}(k d_{1})K_{0}(k d_{2}) - K^{2}_{0}(k d_{1}
    )K^{2}_{0}(k d_{2})\big)$
    The calculation for four-object interaction is not straightforward. In the asymptotic
    limit, $k\to 0$ yields an essential singularity. To incorporate with it, we
    need to consider a cut-off $\Lambda_{\text{cut}}$ sufficiently small but
    different than 0. In asymptotic limit, we can expand the modified Bessel function
    for its large argument as $K_{0}(z) \sim \sqrt{\frac{1}{z}}e^{-z}$. when
    $a \ll \lambda_{D}$

    \begin{equation}
        F^{(4)}\simeq - \frac{5\hbar \omega_{p}a^{8}}{1024 \pi \lambda^{7}_{D}R^{2}}
        \Big[3\, g(\alpha, \beta,R) - X(\alpha, \beta,R) \Big]
    \end{equation}
    when $a \gg \lambda_{D}$
    \begin{equation}
        F^{(4)}\simeq - \frac{5\hbar \omega_{p}a}{1024 \pi }\frac{8\sqrt{2}}{R^{2}\,[\ln(R/a)]^{\frac{7}{2}}}
        \,\Big[ 3\,\Tilde{g}(\alpha, \beta, R) - \Tilde{X}(\alpha, \beta, R)\Big]
    \end{equation}
    where
    \begin{equation}
        g(\alpha, \beta,R) = \frac{\Gamma\Big[0, 2 \Lambda_{\text{cut}}R (1 + \cos(\alpha/2)
        + \cos(\beta/2) \Big]}{2 \sqrt{ \cos(\alpha/2)\cos(\beta/2)}}
    \end{equation}

    \begin{equation}
        \Tilde{g}(\alpha, \beta,R) =\frac{\Gamma\Big[0,
        \frac{2 \Lambda_{\text{cut}}R}{a} (1 + \cos(\alpha/2) + \cos(\beta/2) \Big]}{2\sqrt{
        \cos(\alpha/2)\cos(\beta/2)}}
    \end{equation}

    \begin{equation}
        X(\alpha, \beta,R) = \frac{\Gamma\Big[0, 4 \Lambda_{\text{cut}}R (\cos(\alpha/2)
        + \cos(\beta/2) \Big]}{4 \cos(\alpha/2)\cos(\beta/2)}
    \end{equation}

    \begin{equation}
        \Tilde{X}(\alpha, \beta,R) = \frac{\Gamma\Big[0,
        \frac{4 \Lambda_{\text{cut}}R}{a} (\cos(\alpha/2) + \cos(\beta/2) \Big]}{4\cos(\alpha/2)\cos(\beta/2)}
    \end{equation}
    Here $\Gamma$ is incomplete Gamma function.

    \subsection{Zero-frequency contribution in the free energy}

    \begin{equation}
        \begin{aligned}
            F^{(4)}_{n=0}  \simeq - \frac{k_{B}T a^8}{32 \pi \lambda^{8}_{D}}\Bigg(\int_{\Lambda_{\mathrm{cut}}}^{\infty}dk \, \frac{ \Delta (k) }{\Big[1- \frac{1}{2} (a/\lambda_{D})^2 \ln(ka)\Big]^{4}}\Bigg)
        \end{aligned}
    \end{equation}
where  $\Delta (k) = 3 K^2_{0}(kR) K_{0}(k d_1)K_{0}(k d_2) - K^{2}_{0}(k d_1)K^{2}_{0}(k d_2)$\\ 
    for $a \ll \lambda_{D}$,
    \begin{equation}
        F^{(4)}_{n=0}\simeq - \frac{k_{B}T a^{8}}{32 \pi \lambda^{8}_{D}R}\Big[3\,
        g^{\prime}(\alpha, \beta,R) - X^{\prime}(\alpha, \beta,R) \Big]
    \end{equation}

    \begin{equation}
        \begin{aligned}
            g^{\prime}(\alpha, \beta,R) = \Bigg( \frac{1}{2\Lambda_{\text{cut}}\Big(1 +\cos(\alpha/2)+ \cos(\beta/2)\Big)R}- \\
            \Gamma\Big[0, 2 \Lambda_{\text{cut}}R (1 + \cos(\alpha/2) + \cos(\beta/2) \Big] \Bigg) \times                    \\
            \frac{\Big(1 +\cos(\alpha/2)+ \cos(\beta/2)\Big)}{\sqrt{ \cos(\alpha/2)\cos(\beta/2)}}
        \end{aligned}
    \end{equation}

    \begin{equation}
        \begin{aligned}
            X^{\prime}(\alpha, \beta,R) = \Bigg( \frac{1}{4\Lambda_{\text{cut}} \Big(\cos(\alpha/2)+ \cos(\beta/2)\Big)R}- \\
            \Gamma\Big[0, 4 \Lambda_{\text{cut}}R (\cos(\alpha/2) + \cos(\beta/2) \Big] \Bigg) \times                      \\
            \frac{\Big(\cos(\alpha/2)+ \cos(\beta/2)\Big)}{ \cos(\alpha/2)\cos(\beta/2)}
        \end{aligned}
    \end{equation}

    for $a \gg \lambda_{D}$

    \begin{equation}
      F^{(4)}\simeq - \frac{k_{B}T}{4\pi R\,[\ln(R/a)]^{4}}\,\Big[ 3\,g^{\prime\prime}
        (\alpha, \beta, R) - X^{\prime\prime}(\alpha, \beta, R)\Big]
    \end{equation}
    \begin{equation}
        \begin{aligned}
            g^{\prime\prime}(\alpha, \beta,R) = \Bigg( \frac{1}{2\Lambda_{\text{cut}}\Big(1 +\cos(\alpha/2)+ \cos(\beta/2)\Big)R}- \\
            \Gamma\Big[0, \frac{2 \Lambda_{\text{cut}} R}{a}(1 + \cos(\alpha/2) + \cos(\beta/2) \Big] \Bigg) \times                \\
            \frac{\Big(1 +\cos(\alpha/2)+ \cos(\beta/2)\Big)}{\sqrt{ \cos(\alpha/2)\cos(\beta/2)}}
        \end{aligned}
    \end{equation}

    \begin{equation}
        \begin{aligned}
            X^{\prime\prime}(\alpha, \beta,R) = \Bigg( \frac{1}{4\Lambda_{\text{cut}} \Big(\cos(\alpha/2)+ \cos(\beta/2)\Big)R}- \\
            \Gamma\Big[0, \frac{4 \Lambda_{\text{cut}} R}{a}(\cos(\alpha/2) + \cos(\beta/2) \Big] \Bigg)\times                   \\
            \frac{\Big(\cos(\alpha/2)+ \cos(\beta/2)\Big)}{ \cos(\alpha/2)\cos(\beta/2)}
        \end{aligned}
    \end{equation}


    \section*{Author Contributions}
    Pal, Dobson, and Bostr\"om designed the study. The analytical modelling was done
    by Pal and Dobson supported by Bostr\"om. All authors contributed to the
    writing and overall analysis of the manuscript. All authors have approved the
    final version of the manuscript.

    \section*{Conflicts of interest}
    There are no conflicts of interest to declare.

    \section*{Acknowledgements}
    SP and MB's contributions to this research are part of the project No. 2022/47/P/ST3/01236
    co-funded by the National Science Centre and the European Union's Horizon
    2020 research and innovation programme under the Marie Sk{\l}odowska-Curie
    grant agreement No. 945339.  SP acknowledges support from the Marie Sk{\l}odowska-Curie Doctoral Network TIMES, grant No. 101118915. 
 Institutional and infrastructural support for the ENSEMBLE3 Centre of Excellence was provided through the ENSEMBLE3 project (MAB/2020/14) delivered within the Foundation for Polish Science International Research Agenda Programme and co-financed by the European Regional Development Fund and the Horizon 2020 Teaming for Excellence initiative (Grant Agreement No. 857543), as well as the Ministry of Education and Science initiative "Support for Centres of Excellence in Poland under Horizon 2020" (MEiN/2023/DIR/3797).   
  We gratefully acknowledge
    Poland's high-performance computing infrastructure PLGrid (HPC Centers: ACK Cyfronet
    AGH) for providing computer facilities and support within computational
    grant no. PLG/2023/016228 and for awarding this project access to the LUMI supercomputer,
    owned by the EuroHPC Joint Undertaking, hosted by CSC (Finland) and the LUMI
    consortium through grant no. PLL/2023/4/016319. We thank Profs. Barry W.
    Ninham and Tim Gould for useful discussions.


    \providecommand*{\mcitethebibliography}{\thebibliography}
\csname @ifundefined\endcsname{endmcitethebibliography}
{\let\endmcitethebibliography\endthebibliography}{}

\end{document}